\documentstyle[epsfig]{elsart}

\newcommand{\be}{\begin{equation}}
\newcommand{\ee}{\end{equation}}

\newcommand{\bea}{\begin{eqnarray}}
\newcommand{\eea}{\end{eqnarray}}

\newcommand{\gsim}{\ \rlap{\raise 2pt\hbox{$>$}}{\lower 2pt \hbox{$\sim$}}\ }
\newcommand{\lsim}{\ \rlap{\raise 2pt\hbox{$<$}}{\lower 2pt \hbox{$\sim$}}\ }

\newcommand{\matr}{\left( \begin{array}}
\newcommand{\ematr}{\end{array} \right)}

\makeatletter
\if@twoside
\else \oddsidemargin 00pt \evensidemargin 00pt
\fi
\let\@eqnsel = \hfil

\expandafter\ifx\csname mathrm\endcsname\relax\def\mathrm#1{{\rm #1}}\fi
\@ifundefined{mathrm}{\def\mathrm#1{{\rm #1}}}{\relax}

\makeatother

\begin{document}

\begin{frontmatter}
\title {{\bf  Constraints on neutrino parameters 
and doubly charged Higgs particles 
in the $e^-e^- \rightarrow W^-W^-$ process  }}

\author{ Janusz Gluza}

\address{ Department of Field Theory and Particle Physics, University of
Silesia, Uniwersytecka 4, PL-40-007 Katowice, Poland, 
e-mail address: gluza@us.edu.pl}

\begin{abstract}
Doubly charged Higgs s-channel resonances are analyzed in the $e^-e^-
\rightarrow W^-W^-$ process. In our analysis recent stringent
constraint on an effective neutrino mass $<m_{\nu}>$ coming from
neutrinoless double-$\beta$ decay is implemented.
It is shown that due to a very restrictive limit on $<m_{\nu}>$ 
the
doubly charged Higgs  resonance predicted by the Standard Model with
additional Higgs triplets is below the 
detection limit. For the same reason also the
$\Delta_L^{--}$ resonance can not be visible in the framework of the
conventional Left-Right symmetric model. 
The situation is quite different for the $\Delta_R^{--}$ pole and large
signal is possible here. 
Contributions of the s, t and u channels to the cross section
around the $\Delta_R^{--}$ peak are also discussed.
\end{abstract}
\end{frontmatter}

The $e^-e^- \rightarrow W^-W^-$ process has most recently attracted some 
interest 
\cite{hm1,g1,g2,bel,g3,hm2} as a possible option for the detection of new physics 
in a future 
$\sqrt{s}=0.5-2$ TeV energy $e^-e^-$ facility.
This reaction itself involves many phenomena outside the scope of the Standard
Model (SM). Evidently if the lepton number conservation is violated then even a single 
occurrence
of this process would establish a departure from the SM. The problem of
neutrino mass and its nature is also very closely connected with this process. 
Furthermore, the detection of this reaction could give essential 
information about the 
existence of non-standard doubly charged Higgs particles.

Higgs fields are necessary for the spontaneous
symmetry breaking (SSB) phenomenon. Although in the Standard Model based on the 
$SU_L(2) \times U_Y(1)$ gauge group there is only one physical
scalar Higgs (not observed yet), there is no  restriction for existence of higher
representations of SU(2) for which other physical Higgs particles appear. 
Among them, a triplet representation $(\Delta^{--},\Delta^-,\Delta^0)$ 
that can generate a 
Majorana mass for neutrinos is of special interest \cite{hunt}.
This triplet multiplet includes a doubly charged Higgs field $\Delta^{--}$. 
Its detection would be
the clearest evidence for existence of such a representation in Nature.
Until now we know for sure that  $\Delta^{--}$ with masses below $M_Z/2$
are excluded by the LEPI data \cite{pdg}. However, heavier ones can exist
and the possibility of their detection both in the lepton
\cite{lep1,lep2,lep3,lep4} and
hadron \cite{had1,had2,had3,had4} colliders has already been examined. 
In the $e^+e^-$ and $e^-\gamma$ colliders
$\Delta^{--}$ masses may be probed
almost up to the collision energies \cite{lep4}. At HERA ($e^-p$ collision, 
$\sqrt{s}=313$ GeV) masses up to $m_{\Delta^{--}} \simeq 150$ GeV can be
tested \cite{had1}. Both at the Tevatron \cite{had2} ($p\bar{p}$ collider, 
$\sqrt{s}=2$ TeV) and LHC \cite{had2},\cite{had4} ($pp$ collider, 
$\sqrt{s}=14$ TeV) $\Delta^{--}$ can be found with masses up to 300 GeV
and up to $\simeq 1.2$ TeV, respectively.
So we can see that doubly charged Higgs particles can be found in future
$pp$,
$p\bar{p}$ or $e^+e^-$,$e^-\gamma$ colliders in advance of an $e^-e^-$ option. 
The main feature of
this option is that the $e^-e^-$ initial state is doubly charged and carries
a finite lepton number. Thus SM activity is highly suppressed and clean
signals from any non-standard physics can be looked for there. Especially
direct 
s-channel $\Delta^{--}$ resonances are possible there. After discovering 
this particle in any other facility
the $e^-e^-$ collider could be still used as a $\Delta^{--}$ factory that
could
precisely measure its mass, total decay width and couplings \cite{had2}.
It is also possible that $\Delta^{--}$ will be not found in 
future facilities discussed above so even its approximate mass will be not 
known. Could still the $e^-e^-$ collision be useful in
this case? We will also shortly discussed this situation later on.

s-channel resonances would be established via $\Delta^{--}$ decays. 
Possible
decay modes are $\Delta^{--} \rightarrow W^-W^-,\; \Delta^{--} \rightarrow
\Delta^{-}W^-, \; \Delta^{--} \rightarrow \Delta^{-} \Delta^{-}, \;
\Delta^{--} \rightarrow l^-l^-$. We will focus here on the two gauge 
models
where a non vanishing $\Delta^{--} \rightarrow W^-W^-$ coupling can
exist\footnote{All other possibilities (e.g. when a $\Delta^{--} \rightarrow W^-W^-$ 
coupling is
vanishing) were discussed extensively in \cite{had2} recently.}:

(i) the Standard Model with a Higgs sector
which contains a Higgs-triplet in addition to the standard doublet Higgs fields 
(DTM) \cite{dtm,wud}; 

(ii) the conventional left-right symmetric model (LR) \cite{lr}.

For these models the $e^-e^- \rightarrow W^-W^-$ process with the s channel
exists.

The aim of this work is to investigate this process and to show how its 
$\Delta^{--}$ resonance is connected (and constrained) by neutrinos which
are present in the t and u channels. 
To our best knowledge a discussion of the
$e^-e^- \rightarrow W^-W^-$ process with doubly charged Higgs particles 
has not been performed yet from the point of view of neutrino characteristic
(e.g. their possible mass spectrum, mixing angles with electron and 
CP parities).

In the LR model we have two doubly charged Higgs particles $\Delta_{L}^{--}$
and  $\Delta_{R}^{--}$.

The cross section for s-channel poles ($\sqrt{s}=M_{\Delta_{L,R}^{--}}$)
in the frame of this model can be written in the 
following way ($\beta=\sqrt{1-
4\frac{M_W^2}{s}} $ and  $\pm1/2$ stands for helicities of incoming
electrons)
\begin{equation}
\sigma^{res}(-1/2,-1/2) 
=\frac{4 G_F^2M_W^4\beta}{\pi s} \cos^4{\xi}
\left\{ \left( \frac{s}{2M_W^2}-1 \right)^2 +2 \right\} 
\left[ \frac{\sum_a {(U_L)}_{ae}^2m_a}
{\Gamma_{\Delta_L^{--}} } \right]^2 
,
\end{equation}

\begin{equation}
\sigma^{res}(+1/2,+1/2)  
=\frac{4 G_F^2M_W^4\beta}{\pi s} \sin^4{\xi}
\left\{ \left( \frac{s}{2M_W^2}-1 \right)^2+2 \right\} 
\left[ \frac{\sum_a {(U_R)}_{ae}^2m_a}
{\Gamma_{\Delta_R^{--}} } \right]^2 
.
\end{equation}

Matrices $U_{L,R}$ are a part of the unitary matrix $U={\left( U_L^{\ast},
U_R \right)}^T$ which diagonalize $6 \times 6$ neutrino mass matrix M
(see e.g. \cite{g2})
\begin{equation}
M=\left( \matrix{ M_L & M_D \cr
                M_D^T & M_R} \right),
\end{equation}                

It is important that these s-channel cross sections are proportional to
elements of neutrino mass matrix\footnote{Similarly considering the s channel 
resonance in the $\mu^-\mu^- \rightarrow W^-W^-$ reaction, a future $\mu^-\mu^-$ 
collider could give us information about another two elements of the neutrino 
mass matrix M, e.g. ${(M_L^{\ast})}_{\mu\mu}$ and ${(M_R)}_{\mu\mu}$.}
\begin{eqnarray}
\sum_{a} \left( U_L^T \right)_{ea}m_a \left( U_L \right)_{ae}&=&\left( 
M_L^{\ast} \right)_{ee}, \nonumber \\
\sum_{a} \left( U_R^T \right)_{ea}m_a \left( U_R \right)_{ae}&=&\left( 
M_R \right)_{ee}.
\end{eqnarray}
The situation is akin to that in the ${(\beta\beta)}_{0\nu}$
process. It has been shown that for any realistic local gauge theory the
existence of ${(\beta\beta)}_{0\nu}$ process implies that (an effective) 
electron neutrino must be of Majorana type with a nonzero mass \cite{val}. 
For the $e^-e^- \rightarrow W^-W^-$
process we can also recall another argument.
A proper high energy behaviour of the cross section 
\cite{g1},\cite{orig} demands that all three s,t and u channels are needed
simultaneously.
That means that massive neutrinos must exist and they are exchanged in t and 
u channels\footnote{Massless neutrinos give
vanishing contributions of t and u channels to the total cross section
even when right-handed interactions exist \cite{g1}}. 

The $\xi$ angle is connected with diagonalization of the charged gauge 
boson mass matrix and can be expressed in the following way \cite{g1}
\begin{equation}
\xi \simeq  \frac{M_{W_1}^2}{M_{W_2}^2}.
\end{equation}

The full cross section with t and u channels will not be given here as it 
can be found in e.g. \cite{g1},\cite{g2}. The only important 
information
we need to know is that the helicity amplitudes for t and u channels with
exchanged neutrinos are not
directly proportional to ${(M_{L,R})}_{ee}$ (Eqs.(1-3)) because an
additional neutrino mass relation comes from the neutrino propagator.
However, for energies where the
s-channel resonance appears these channels are negligible.
Outside the s-channel resonance they can be important and then both 
mixing matrix elements ${(U_{L,R})}_
{ee}$ and masses $m_a$ must all be known separately.

In the frame of the TDM model we are restricted to a $3 \times 3$ neutrino
mass matrix $M=M_L$ \cite{wud} and the cross section for the $\Delta_L^{--}$
pole can be obtained from Eq.(1) by putting $\xi=0$ (that means that only
left-handed currents exist).

Let's proceed to numerical results. 

The DTM  model predicts three massive neutrinos which can be identified with 
electron, muon and tau, respectively. They enter to the formula
on $\sigma^{res}$ by the 
$\sum_{a} \left( U_L \right)^2_{ea}m_a$ factor.
However, this quantity is directly constrained by the absence of the 
$\left( \beta\beta \right)_
{0 \nu}$ decay. From experimental data an effective electron
neutrino mass is extracted to be \cite{light} (modified to our notation)
\begin{equation}
<m_{\nu}>=(M_L^{\ast})_{ee} =\sum_{a} \left( U_L \right)^2_{ea}m_a \leq 0.65\; \mbox{\rm eV},
\end{equation}
which implies that we don't need to consider masses of all neutrinos
and strength of their couplings with electron 
separately. To show how extremely strong this limit is let's take the 
$\Delta_L^{--}$ resonance at CM energy equal to 200 GeV (
$\sqrt{s}=200$ GeV = $M_{\Delta_L^{--}}$) and $\Gamma_{\Delta_L^{--}}
\simeq 10$ GeV. Value of the width was adopted from \cite{wud} where 
$\Delta_L^{--}$
Higgs' lifetime has been analyzed. This value varies with $s_H$
($s_H=\sqrt{8v_T^2/(8v_T^2+v_D^2)}$, $v_{T,D}$ are vacuum expectation values
for Higgs triplet and doublet fields, respectively)
about one order of magnitude for $0.1<s_H<0.995$.

The cross section is then
\begin{equation}
\sigma^{res(\Delta_L^{--})} \left( \sqrt{s}=200\;GeV \right) \simeq 10^{-14}\;fb.
\end{equation}
Unavoidably it is below detection level. 

For larger energies (and larger masses of doubly charged Higgs particles)
the total width increases and the cross section at the $\Delta_L^{--}$ pole is 
smaller than given in Eq.(7). On the other hand the resonance would have 
to be
extremely narrow ($\Gamma_{\Delta_L^{--}} \sim 0(1)$ keV) if the signal
were to be detectable ($\sigma^{res(\Delta_L^{--})} \sim 0(1)$ fb).

We can conclude that the $\Delta_L^{--}$ resonance from the DTM model is under 
detection.

There are two doubly charged Higgses ($\Delta_{L,R}^{--}$) in 
the LR model.
From Eqs.(2) and (6) we infer that situation 
for the $\Delta_L^{--}$ resonance is quantitatively
the same as in the DTM case.

However, situation is different for the $\Delta_R^{--}$ resonance (Eq.(2)) because 
$\sigma^{res}(\Delta_R^{--})$ is proportional to 
$\sum_{a} (U_R)_{ae}^2m_a =(M_R)_{ee}$ (Eq.(4)).
Taking into account that ${(M_R)}_{lk} >> {(M_D)}_{lk}$ and  $M_R 
\leq
\frac{2M_{W_2}}{g}$ \cite{g1} we have
\begin{equation}
O(1)\;{\rm GeV} << (M_R)_{ee} \leq \frac{2M_{W_2}}{g}.
\end{equation}
We can see (Eq.(2)) that this factor can greatly enhance the 
resonance signal. Note, however, that this happens for +1/2 helicity
polarization of incoming electrons where the reduction factor $\sin^4{\xi}$
is present so both factors must be examined simultaneously more carefully.

Let's consider two versions of the left-right symmetric model, so-called 
Manifest or 
Quasi-Manifest L-R symmetric model (MLRS) and Non-Manifest L-R symmetric model 
(NMLRS). In the frame of NMLRS the present experimental bound on $M_{W_2}$
is not so high and $M_{W_2} \geq $600 GeV is still possible \cite{mwa}.
However, for
MLRS models the bound is larger, $M_{W_2} \geq$ 1600 GeV \cite{mwb}.
For these models in numerical discussion we use the lower limits.

Fig.1 portrays results for the $\Delta_R^{--}$ resonance 
($\sqrt{s}=M_{\Delta_R^{--}}$) for the NMLRS model. All lines are
designed to maintain constant $\sigma^{res(\Delta_R^{--})}=1$ pb
while changing ${(M_R)}_{ee}$ and 
$\Gamma_{\Delta_{R}^{--}}$. Values of ${(M_R)}_{ee}$ 
satisfy inequality (8).

Above these lines (larger $\Gamma_{\Delta_R^{--}}$) $\sigma^{res} <$1 pb.

We can see from this Figure that the reference level 
$\sigma^{res(\Delta_R^{--})}=1$ pb
(i.e. about $10^4$ times larger than `the detection limit' for this process 
\cite{kol})
can be easily achieved for $\sqrt{s} \geq 500$ GeV for a wide range of 
${(M_R)}_{ee}$ and $\Gamma_{\Delta_{R}^{--}}$ parameters.
For $M_{W_2}$=1600 GeV (the MLR model) $\sin{\xi}$ (Eq.(5)) is smaller and to achieve
the same values of $\sigma^{res}(\Delta_R^{--})$ much smaller values
of $\Gamma_{\Delta_R^{--}}$ are needed (e.g. $\Gamma_{\Delta_R^{--}} \leq 
O(10)$ GeV for $\sqrt{s}=1$ TeV).

Fig.1 shows only optimistic, on-peak results. A question may arise how
this situation looks like at the off-peak s-channel energy region
where s-channel contribution is less pronounced and t and u channel contributions
start to play a role. This case is discussed in Fig.2 for 
$M_{\Delta_R^{--}}=500$ GeV with two parameters taken from the dotted line in 
Fig.1, i.e. $\Gamma_{\Delta_R^{--}} \simeq 10$ GeV and ${(M_R)}_{ee}=500$ GeV
(the asterisk).
As it was already mentioned to make numerical calculations pertaining to off-peak
energies we need 
to know not only the value of ${(M_R)}_{ee}$ itself, as in the on-peak case
(Eq.(2)), but also mixing matrix elements
between heavy neutrinos and electron ${(U_{L,R})}_{eN_i}$ 
(i=1,2,3). It can be shown that  the
mixing matrix element ${(U_{R})}_{eN}$ can be very large, ${(U_{R})}_{eN}\simeq
1$ (see \cite{g4} for theoretical and \cite{g5} for practical realization of this situation).
Then, because of unitarity of the U mixing matrix other elements must 
fulfill the relation (i=2,3)
\begin{equation}
|{(U_{R})}_{eN_i}|<< |{(U_{R})}_{eN}|.
\end{equation}

Taking into account relevant experimental constraints on heavy neutrino
mixing angles with electron \cite{light},\cite{tom} we can deduce (see 
\cite{g3},\cite{g6} for details) that the maximal mixing between 
heavy neutrino and electron predicted by present data and the LR model
is ${(U_L)}_{Ne}^2 \approx 0.0027$.
In  computations we take ${(U_L)}_{Ne}^2$=0.001.

As we can see from Fig.2 for the discussed left-right model parameters, the  
contribution of the s channel to the total 
cross section is meaningful also outside the s channel resonance (i.e. for
$\sqrt{s}=500\pm 10 \Gamma_{\Delta_{R}^{--}}$). That means that also without
knowledge about the mass of $\Delta_R^{--}$ (which can originate from other
facilities) we can try to look for it by fixing
as many as possible energy settings to cover a wide range of possible
$\Delta_R^{--}$ masses\footnote{This statement can be true only for large
decay width of $\Delta_R^{--}$ 
(i.e. for $\Gamma_{\Delta_{R}^{--}} \geq 1$ GeV). The Gaussian
shape of the $\sqrt{s}$ spectrum depends on beamstrahlung/bremsstrahlung
effects. For $\sqrt{s}=500$ GeV an average energy loss from these effects is
estimated to be of the order of 3\% and a Gaussian peak can be controlled given
pretty nice luminosity $L=50\;fb^{-1}$ for a few years of running in its central
0.2\%  peak region  
(i.e. for the Gaussian rms resolution $\sigma$ equals 1 GeV) \cite{had2}. For
$\Gamma_{\Delta_{R}^{--}} \leq \sigma$ information
about the mass of $\Delta_R^{--}$ particle 
would be desirable. Otherwise we could measure a signal not at the central peak
but at a tail of the $\sqrt{s}$ spectrum where the luminosity would be not
sufficient for getting an observable signal. For more details, among other
things about predicted luminosities and beamstrahlung effects in the
$\sqrt{s}=0.5 \div 2$ TeV $e^-e^-$ collisions, see e.g. \cite{exp}.}. 
Similar plots can be obtained also for both larger and smaller energies,
as long as considered energies are at or nearby the $M_{\Delta_R^{--}}$ pole
($\sqrt{s}=M_{\Delta_R^{--}}\pm \mbox{\rm (few)}\; \Gamma_{\Delta_{R}^{--}}$) and mixing
angle $\xi$ is large enough (e.g. $M_{W_2}$ is not too massive (Eq.(5)).


Finally, let's note that as long as the s channel dominates over t and u 
channels we can hardly 
say anything about CP parities of heavy neutrinos (the mixing angle of only
one heavy neutrino with electron is important (Eq.(9)) and then 
interferences with the other ones disappear \cite{g1}).

In conclusion, we have shown, that due to a very precise bound on the effective
neutrino mass $<m_{\nu}>$ taken from $(\beta\beta)_{0\nu}$ experiment,
the s-channel resonance  predicted by the DTM model is below
detection in the $e^-e^- \rightarrow W^-W^-$ process. 
The LR model opens a possibility for its detection, especially 
when the NMLR version of this model (with a possible small value of the additional
charged gauge boson $M_{W_2}$) is considered.
Then the $\Delta_R^{--}$ Higgs particle resonance is attainable
for a wide range of doubly charged Higgs masses and total widths. 
Detecting this resonance would give also information that
massive Majorana neutrinos exist.

\newpage

\section*{Acknowledgments}

I'd like to thank prof. M.~Zra\l ek for careful reading and valuable remarks 
on this letter.

This work was supported by the Polish Committee for Scientific Research
under Grant No. PB659/P03/95/08 and an internal Grant of the 
University of Silesia. I also appreciate the fellowship of the Foundation
for Polish Science.


\newpage
\ \\

\vspace{7.5 cm}
\begin{figure}[h]
\includegraphics{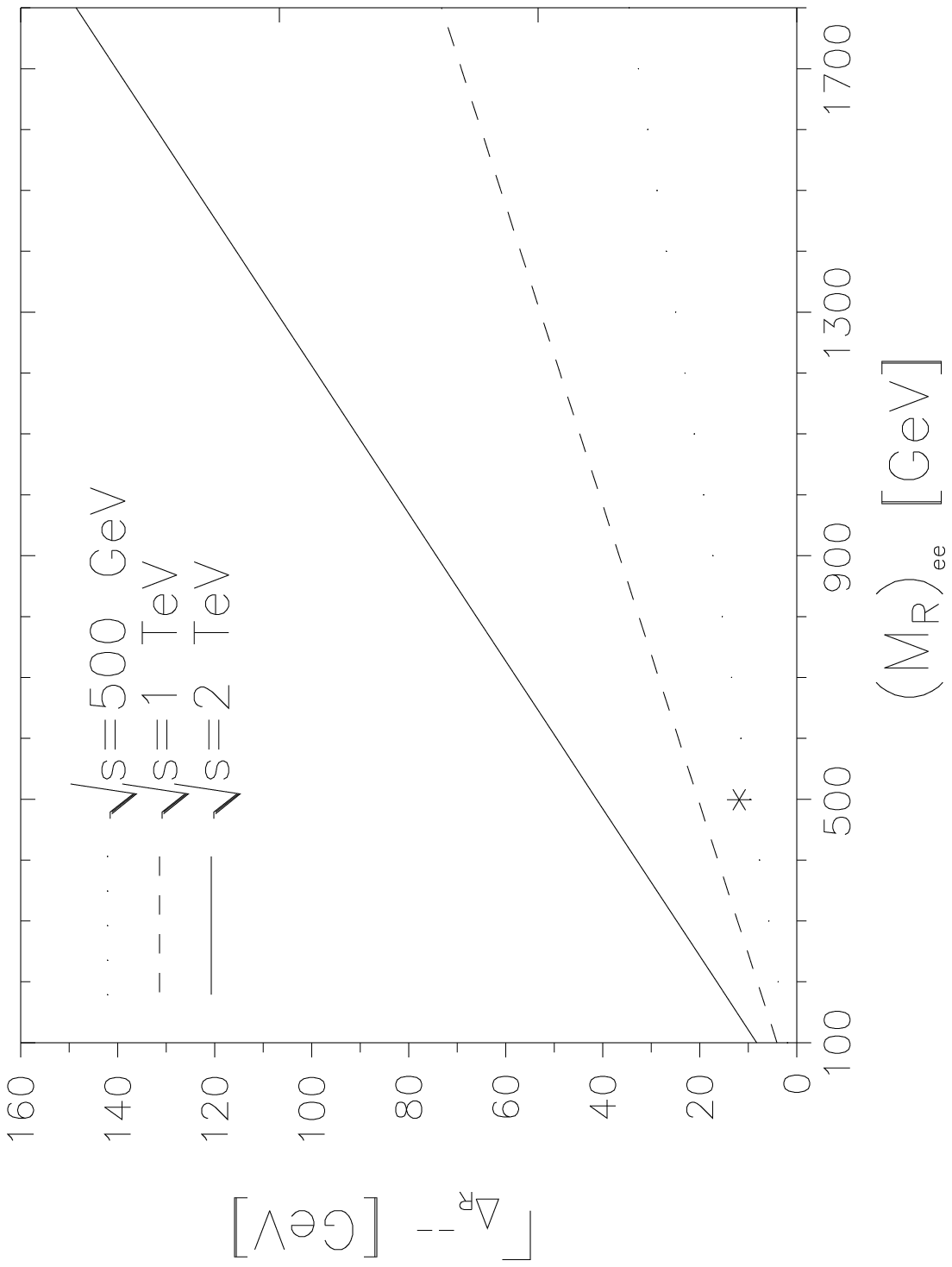}
\vspace{0.3 cm}
\end{figure}

\baselineskip 5 mm

{\footnotesize Fig.1 
This Figure shows lines with $\Gamma_{\Delta_R^{--}}$ and ${(M_R)}_{ee}$
parameters for which  $\sigma^{res}(\sqrt{s}=M_{\Delta_R^{--}}) 
=1$ pb. Solid, dashed and
dotted lines are for $\sqrt{s}=M_{\Delta_R^{--}}=$2000,1000,500 
GeV, respectively. Above these lines $\sigma^{res}
(\sqrt{s}=M_{\Delta_R^{--}}) < 1$ pb. The asterisk denotes a point in
$\Gamma_{\Delta_R^{--}}-
{(M_R)}_{ee}$ coordinates which is used as a parameter in further 
calculations (Fig.2).}

\vspace{0.5 cm}

\vspace{7.5 cm}
\begin{figure}[h]
\includegraphics{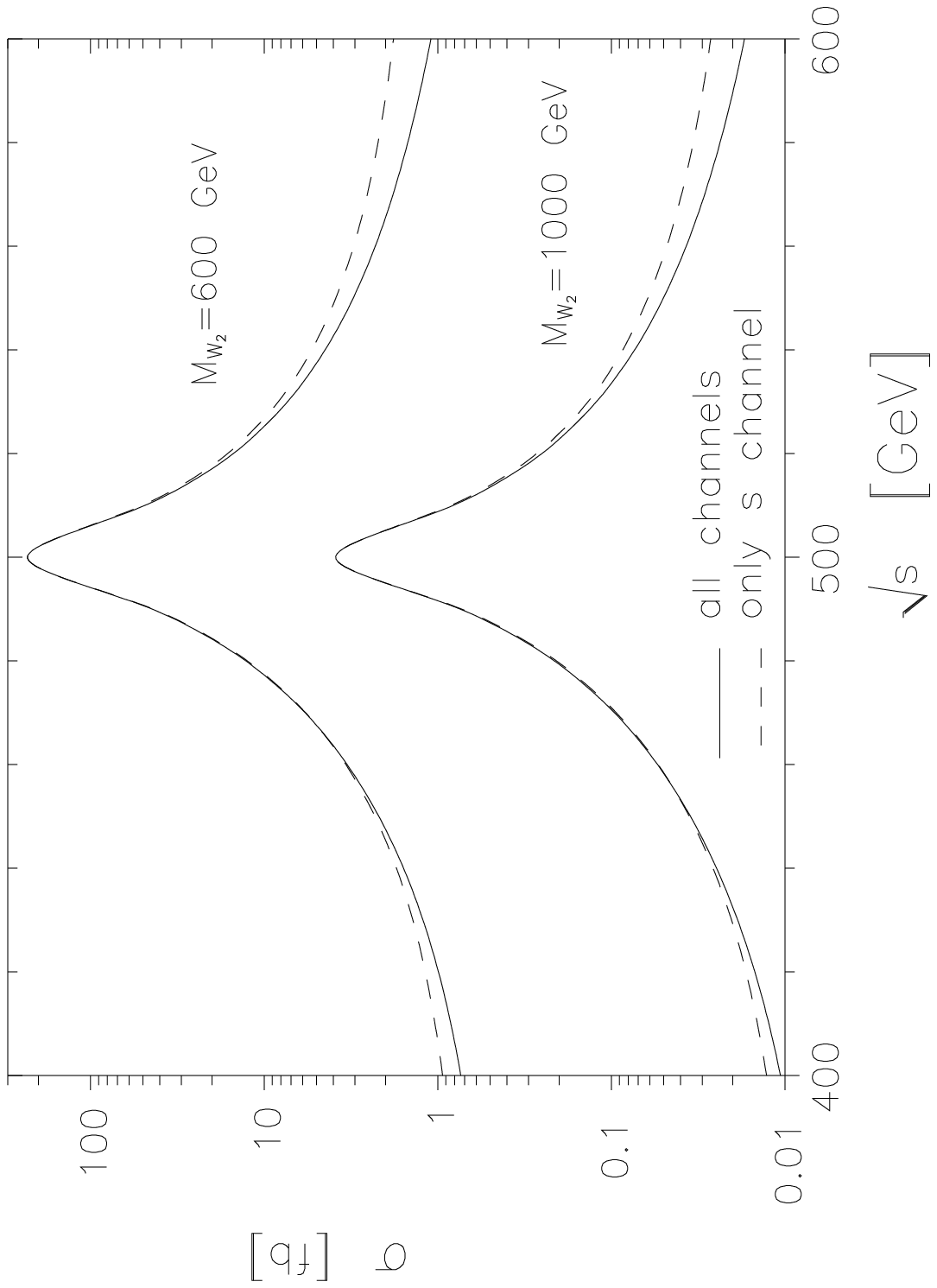}
\vspace{0.3 cm}
\end{figure}

\baselineskip 5 mm

{\footnotesize Fig.2 The cross section for the $e^-e^- \rightarrow W^-W^-$ process as a function
of energy near the s-resonance ($M_{\Delta_R^{--}}=500$ GeV) for
the left-right symmetric model with $\Gamma_{\Delta_R^{--}}=10$ GeV
and ${(M_R)}_{ee}=500$ GeV (the asterisk in Fig.1). Solid lines stand 
for the total cross sections,
dashed lines stand for the cross sections after removing t and u channels
(the s channel contribution only).}


\begin{thebibliography}{99}
\bibitem{hm1} C.A.~Heusch and P.~Minkowski, Nucl.Phys.{\bf B416}(1994)3.
\bibitem{g1} J.~Gluza and M.~Zra{\l}ek, Phys.Rev.{\bf D52}(1995)6238.
\bibitem{g2} J.~Gluza and M.~Zra{\l}ek, Phys.~Lett.{\bf B362}(1995)148.
\bibitem{bel} G.~Belanger, F.~Boudjema, D.~London and H.~Nadeau,
Phys.Rev. {\bf D53}(1996)6292; T.~G.~Rizzo, Int.J.Mod.Phys. {\bf
A11}(1996)1613.
\bibitem{g3} J.~Gluza and M.~Zra{\l}ek, Phys.~Lett.{\bf B}372(1996)259. 
\bibitem{hm2}C.A.~Heusch and P.~Minkowski,hep-ph/9611353.
\bibitem{hunt} J.~Gunion, H.~Haber, G.~Kane and S.~Dawson `The Higgs Hunters Guide'
(Addison-Wesley, Menlo Park, 1990).
\bibitem{pdg} P.D.~Acton et al., Phys.Lett.{\bf B295}(1992)347.
\bibitem{lep1} M.~Lusignoli and S.~Petrarca, Phys.Lett. {\bf B226}(1989)397;
\bibitem{lep2} M.L.~Swartz, Phys.Rev. {\bf D40}(1989)1521;
R.~Godbole, B.~Mukhopadhyaya, M.~Nowakowski, Phys.Lett.
{\bf B352}(1995)388; V.~Barger, J.F.~Beacom, K.~Cheung and T.~Han, 
hep-ph/9505335;
J.A.~Coarasa, A.~Mendez and J.~Sola, Phys.Lett.{\bf B374}(1996)131;
\bibitem{lep3} J.F.~Gunion, Int.J.Mod.Phys.{\bf A11}(1996)1551
\bibitem{lep4} G.~Barenboim, K.~Huitu, J.~Maalampi and M.~Raidal, 
hep-ph/9611362.
\bibitem{had1} E.~Accomando, S.~Petrarca, Phys.Lett. {\bf B323}(1994)212;
\bibitem{had2} J.F.~Gunion, C.~Loomis, K.T.~Pitts, hep-ph/9610237.
\bibitem{had3} H.~Georgi, M.~Machacek, Nucl.Phys. {\bf B262}(1985)463;
R.~Vega, D.A.~Dicus, Nucl.Phys. {\bf B329}(1990)533;
E.~Accomando, M.~Iori, M.~Mattioli, hep-ph/9505274; 
\bibitem{had4}K.~Huitu, J.~Maalampi, A.~Pietila and M.~Raidal,
hep-ph/9606311.
\bibitem{dtm} H.M.~Georgi, S.L.~Glashow and S.~Nusinov, 
Nucl.Phys.{\bf B193}(1981)297;
S.~Chanowitz and M.~Golden, Phys.Lett.{\bf B165}(1985)105.
\bibitem{wud} J.F.~Gunion, R.~Vega and J.~Wudka, Phys.Rev.{\bf D42}(1990)1673.
\bibitem{lr} J.C.~Pati and A.~Salam, Phys.Rev.{\bf D10}(1974)275;
R.N.~Mohapatra and J.C.~Pati, ibid. {\bf D11}
(1975)566; {\bf D11}(1975)2559; G.~Senjanovic and R.N.~Mohapatra, ibid. 
{\bf D12}(1975)152; G.~Senjanovic, Nucl.Phys.{\bf B153}(1979)334.
\bibitem{val} J.~Schechter and J.W.F.~Valle, Phys.Rev.{\bf D25}(1982)2951.
\bibitem{orig} T.G.~Rizzo, Phys.Lett. {\bf B116} (1982)23.
\bibitem{light} A.~Balysh et.al. (Heidelberg-Moscow Coll.), Phys.Lett.
{\bf B356}(1995)450. 
\bibitem{mwa} F.~Abe et.al., Phys.Rev.Lett.67(1991)2609.
\bibitem{mwb} G.~Beall, M.~Bander and A.~Soni, Phys.Rev.Lett.48(1982)848.
\bibitem{kol} J.F.~Gunion and A.~Tofighi-Niaki, Phys.Rev. {\bf D36}(1987)2671; 
{\bf D38}(1988)1433; T.G.~Rizzo, Waikoloa Linear Collid.(1993)520;
F.~Cuypers, K.~Ko{\l}odziej, O.~Korakianitis and R.~Ruckl, Phys.Lett. {\bf
B325}(1994)243.
\bibitem{g4} J.~Gluza and M.~Zra\l ek, Phys. Rev.{\bf D51}(1995)4695.
\bibitem{g5} J.~Gluza and M.~Zra\l ek, Phys. Rev.{\bf D48}(1993)5093.
\bibitem{tom} E.~Nardi, E.~Roulet and D.~Tommasini, Nucl.Phys. {\bf B386}
(1992)239; A.~Djouadi, J.~Ng., T.G.~Rizzo, `New particles and interactions',
hep-ph/9504210; T.~Bernatowicz et al., Phys.Rev.Lett. {\bf 69}(1992)2341;
M.~Hirsch, H.V.Klapdor-Kleingrothaus, O.~Panela, Phys.Lett. {\bf
B374}(1996)7.
\bibitem{g6} J.~Gluza, M.~Zra{\l}ek, hep-ph/9612227;
J.~Gluza, J.~Maalampi, M.~Raidal and M.~Zra\l ek, in preparation.
\bibitem{exp} T.W.~Markiewicz, Int.J.Mod.Phys. {\bf A11}(1996)1649;
B.~A.~Schumm, ibid. 1667; J.~E.Spencer, ibid. 1675; Pisin Chen et al.,
ibid. 1687; R.~Erickson, ibid. 1693.
\end{thebibliography}
\end{document}